# Fundamental limits of hot carrier injection from metal in nano plasmonics

*Jacob B. Khurgin*

*Johns Hopkins University*

*Baltimore MD 21218 USA*

Evolution of the non-equilibrium carriers excited in the process of decay of surface plasmon polaritons (SPPs) in metal is described for each step – from the carrier generation to their extraction from the metal. The relative importance of various carrier generating mechanism is discussed. It is shown that both carrier generation and their decay are inherently quantum processes as for realistic illumination conditions no more than a single SPP per nanoparticle exists at a given time. As a result, the distribution of non-equilibrium carriers cannot be described by a single temperature. It is also shown that the originally excited carriers that have not undergone a single electron-electron scattering event, are practically the only ones that contribute to the injection. The role of the momentum conservation in the carrier extraction is discussed and it is shown that if all the momentum conservation rules are relaxed, it is the density of states in the semiconductor/dielectric that determines the ultimate injection efficiency. A set of recommendations aimed at improving the efficiency of plasmonic-assisted photodetection and (to a lesser degree) photocatalysis is made in the end.

## 1. Introduction

The last two decades have seen vigorous growth of exploration in plasmonics [1-3]and, in a broader sense, in the interaction between the light and free carriers in the metal or other media (such as doped semiconductors). The salient feature of metallic structures is their ability to concentrate optical fields into the small volumes that are not limited by diffraction. A large number of plasmonic devices with enhanced performance in various regions of electro-magnetic spectra has been conceived [4] and to a certain degree demonstrated since the turn of the millennium, including sources [5], detectors[6, 7] and modulators[8, 9] of radiation, as well as wide range of sensors. Yet, with exception of sensors[10, 11], plasmonic devices have failed to enter the mainstream for one reason – very high loss inherent to all metals. This loss is innate to metals because large fraction of energy is contained in kinetic motion of carriers which scatter the energy at the rate of up to $10^{14} s^{-1}$ or even more [12]. Or, if one wants to put it in a "quantum mindset", the aforementioned large loss can be explicated by the large density of both occupied and empty states below and above Fermi energy respectively and hence large rate of transitions between those states as postulated by the Fermi Golden rule. The loss is particularly high at shorter wavelengths while at longer wavelength the damage from loss is less extensive [13-15] as the electromagnetic field does not really penetrate the metal, so the functional metal-based devices in the mid-IR and THz range are not really plasmonic in the correct sense of that word.

It is the realization that loss in plasmonics is inevitable that has prompted a significant part of plasmonic community to re-examine the issue and shift the focus of their efforts from the futile battle to reduce the absorption in metals to the quest for creative use of that absorption [16, 17], which in fact should not be thought of as an irretrievable loss but rather as the transfer of energy from plasmons first to the single particle excitations in metal (electron-hole pairs) and then to the lattice vibrations. If the absorbed energy can be recaptured on one of the stages before the equilibrium with the surroundings is achieved, it can be put into productive use as has been indeed demonstrated by a number of groups[6, 18-20]. The stage at which the absorbed energy can be captured with the least effort is obviously just after the energy has been transferred to the lattice (which in general is far less than a picosecond). Depending on how well the plasmonic entities are isolated from the surroundings, they can be heated by hundreds of degrees and that rise in temperature can be used for diverse applications ranging from cancer therapy [21] to thermophotovoltaics[22].

Capturing the energy at the first stage, when it has just been transferred from the plasmon polaritons to the electron-hole pairs and before it has moved down the line to the lattice phonons is obviously significantly more difficult since it has to be done on sub-picosecond scale, but is also potentially far more rewarding since the kinetic energies (relative to the Fermi energy) of these so-called "hot carriers" are commensurate with the photon energy, i.e. correspond to tens of thousands of degrees of Kelvin, i.e. way beyond the melting point of the metals. For this reason, the hot carriers have sufficient energy to perform the feats that carriers in equilibrium with the lattice can never do, even if the lattice is heated almost to the point of melting. In particular, hot carriers may have energy sufficient to overcome the binding force that keeps them inside the metal and carrying charge out and into the adjacent material which can be semiconductor, dielectric, or a solution surrounding the metal nanoparticle. Once the barrier is surmounted, the charges are separated, and by collecting the charges at contacts, one can operate the device as a photodetector or a photovoltaic cell as has indeed been done in many works. Alternatively, the charges can serve as catalysts for various chemical reactions that take place either directly on the metal surface [23-25], or, more often after first being transferred to a dielectric or semiconductor material, like $TiO_2$[26, 27].

At this time a large body of work has been accumulated on the process of charge transfer of hot carriers from the plasmonic metals into semiconductors or dielectrics [27-38]. The results (in terms of the injection efficiency) have varied between the different groups, sometimes by orders of magnitude and more. The confusion has not been greatly helped by numerous theories that have been developed to explain the observed phenomena. Trying not to sound too critical of all the worthwhile efforts expended in the attempt to make sense of hot carrier injection, we would nevertheless mention the common trend of overreliance on the numerical methods of many of the prior works to the detriment of physical understanding. Moreover, recently the whole concept of carrier injection has been put into question and an attempt had been made to explain many experimental data simply via the heating of nanoparticles (i.e. what we refereed before to as a second stage of energy transfer) [39, 40]. As shown later this valiant endeavor, while not being entirely correct, is definitely not without a merit.

Faced with this reign of confusion in the field that I undertake this modest effort to shed some light onto hot carrier genesis, exodus, and decay that plagues the latter in plasmonic metals. As I already mentioned, this is not an homage to all the prior works, nor is it their explicit critique, so the readers glancing through these pages with the sole goal of establishing that a given prior work has been mentioned (no matter in what context), will be disappointed and should seek and find solace in multiple articles proliferating elsewhere. In my opinion, referring the reader to a few solid review articles[41, 42] in addition to all mentioned above is sufficient to establish the framework for the present work. The goals of this work are very modest – to present what I believe is a simple physical picture of hot carrier generation, decay and emission and to establish what can be a maximum injection efficiency for a given metal/semiconductor (dielectric) combination.

The work is structured in the following way. Section 2, understandably, if ambitiously, entitled "Genesis", is essentially a very short review of my prior work [43] that has established how the hot carriers are generated by different mechanisms and what is their distribution in energy, space and angular coordinates. But also in this section an often overlooked feature in plasmonics is uncovered – the fact that under most practical conditions only a single SPP is excited on a given nanoparticle. In Section 3, that in line with the Pentateuch is called "Numbers", I make a key distinction between the "ballistic" or "first generation" carriers produced by light and all the subsequent "generations" spawned by the fast electron-electron scattering, with progressively lower energies, all the way to the thermalized hot carriers that can be characterized by the electron temperature $T_e$. I show that for the most practical situations, it is only that first "generation" of ballistic carriers that stands a decent chance of surmounting the barrier keeping them inside the metal. In Section 4 logically called "Exodus", I consider the role of lateral momentum conservation in the electron transfer across the barrier and show that if momentum conservation is fully relaxed by the interface roughness the injection efficiency can be greatly increased and depends only on the density of states (DOS) in two adjacent media. It this DOS ratio that determines the ultimate limit for electron injection from the metal, in total analogy to the "4n²" limit for the light capture in dielectric determined by Yablonovitch almost four decades ago [44, 45]. Section 5 is unsurprisingly devoted to the conclusions that hopefully will be of some use to the community.

## 2 Genesis: Hot carriers are generated. How and where?

### 2.a How many SPPs are excited simultaneously?

Before reviewing the mechanisms leading to SPP decay into electron hole pairs, it is instructive to ascertain roughly the numbers (for localized) and densities (for propagating) of SPPs involved in this. One can start with the localized SPPs as shown in Fig.1 a in which the electric field $E_{SPP}$ is enhanced relative to the incident field $E_{IN}$ by roughly a factor of $F \approx Q = \omega / \gamma$ [1, 46], where $\gamma$ is the total SPP decay rate (radiative and nonradiative) to be evaluated later. This ratio can be higher or lower depending on the geometry and if resonant nano-antennas are involved the ratio can be as high as $F \sim Q^2$ [47, 48]. Therefore the energy density in the SPP mode can be found as $u_{SPP} = F^2 I_{IN} n / c$ where $I_{IN}$ is the power density of the incoming radiation and n is the index of refraction surrounding the nanoparticle. The total energy of the SPP mode is then $U_{SPP} = F^2 I_{IN} n / c \times V_{SPP}$ where $V_{SPP}$ is a volume of the SPP mode, commensurate with the size of nanoparticle. Consider then a spherical nanoparticle of 50nm diameter and the light with average wavelength of 500nm and refractive index n=1. Then the average number of SPPs in the mode is $N_{SPP} = U_{SPP} / \hbar \omega \approx 2 \times 10^{-9} F^2 I_{IN}$, where incident power density is in $W/cm^2$. Now even if we assume a really high field enhancement factor $F^2 = 10^3$ one can see that in order to have on average one SPP at a given time the incoming radiation should have power density of $1 MW/cm^2$ in a spectral region around SPP resonance. For comparison, solar irradiance near equator (1 sun) is only $I_{IN} \sim 0.1 W/cm^2$ so even concentrating this power by a factor $10^3$ would still leave us with far less than 1 SPP at any given time. The situation is shown schematically in Fig. 2(a) where among many nanoparticles only a single one marked as "A" supports an SPP. After a plasmon lifetime $\tau_{SPP} = \gamma^{-1}$ elapses there are no SPPs left within the volume as shown in Fig.2b while the SPP energy is transferred to hot carriers in the marked nanoparticle "A". Then (Fig.2c) after another $\tau_{SPP}$ (although it can take significantly longer) an SPP is excited on the nanoparticle "B", and so on. Obviously, in photodetectors the irradiance is typically far less than $1 MW/cm^2$ as well. In other words, in practical situations, most of the time a given nanoparticle has no SPP in it –a single SPP is excited on rare occasions and then very quickly decays. That indicates that the SPP decay is by definition a quantum process that cannot be described classically as only one SPP is there at any given time. Another way to look at it is to establish the average interval between SPP excitations of a given nanoparticles as $\Delta\tau = \tau_{SPP}/N_{SPP}$ which for the 50nm nanoparticle with $\tau_{SPP} = 10 fs$ considered above $\Delta\tau = 0.5\mu s/I_{IN}$. So it takes microseconds for the same nanoparticle gets excited again! Hence the situation depicted in Fig.1a in which the SPP mode is excited while some electron-hole pairs are present in the nanoparticle is not realistic – at a given time there may be an SPP excited on a nanoparticle or a hot electron-hole pair in it, or, most probably no excitation at all.

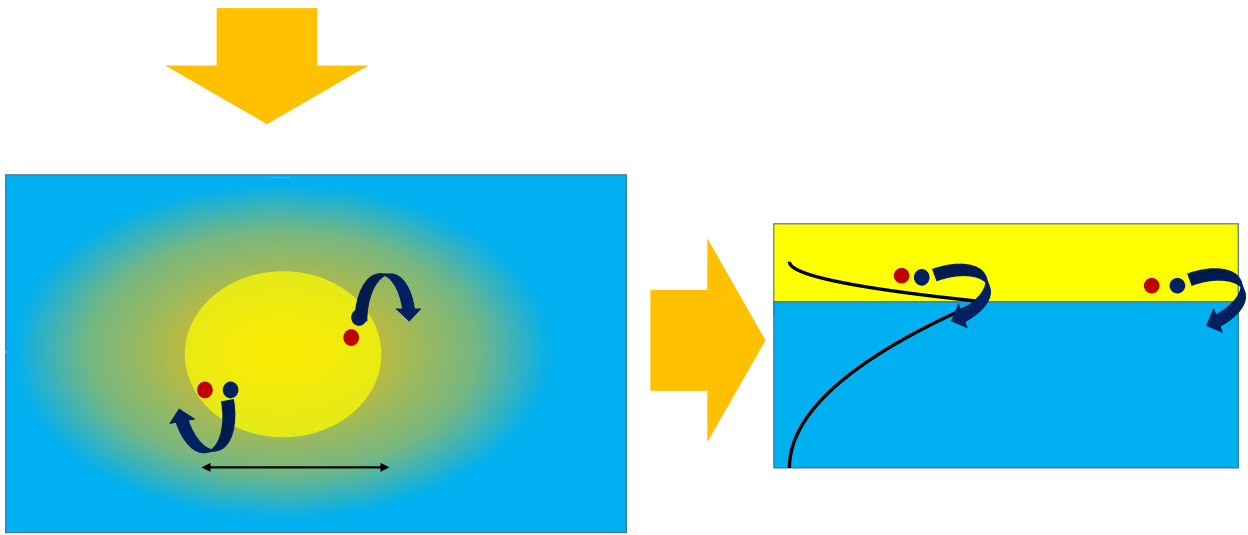

**Fig.1** Geometry of hot electron-hole pair generation in the metal with subsequent injection into semiconductor/dielectric using (a) Localized SPP's (mostly used in photocatalysis) and (b) Propagating SPPs (mostly used in photodetection). At realistic optical powers SPPs never coexist with hot carriers at the same time.

In the other relevant geometry (Fig.1b), if we consider a plasmonic or hybrid waveguide in which $P_{IN} = 1\mu W$ of power (which is a lot for the detector!) propagate while being absorbed in metal, the linear density of the propagating SPP's is $dN_{SPP}/dl = P_{in}n/c\hbar\omega \approx 0.01/\mu m$. Since typically absorption takes place over the distance of only a few micrometers, once again the average number of SPP's inside the plasmonic waveguide at a given time is less than one, and, once again the process of hot carrier excitation should be treated as a quantum process.

It should be noted that in many experimental works the SPP excitation and hot carrier regeneration and decay have been studied using femtosecond lasers [29, 49-51] in which power densities can be much higher than a MW/cm$^2$. These time-resolved measurements do provide information about hot carriers evolution, however they do it for the situation of multiple hot carrier excitation which actually almost never occurs in practical applications. For instance, when multiple SPPs are generated on the same nanoparticle there is always a possibility of exciting the carriers with energy exceeding $\hbar\omega$, but when the interval between SPP generation is long such even cannot take place.

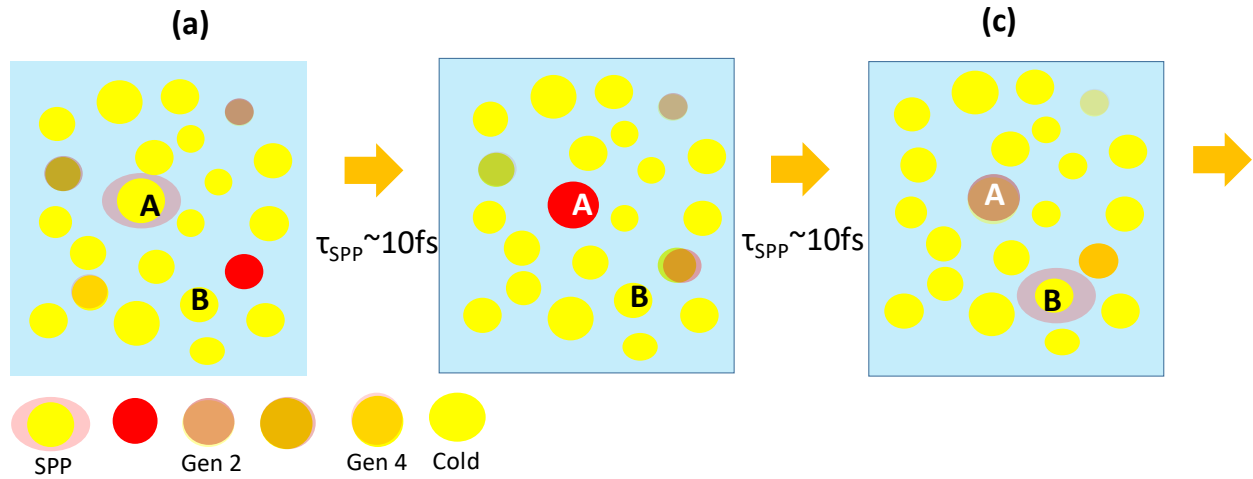

**Fig.2** Dynamics of hot carrier generation and decay in the array of metal nanoparticles under typical photo-excitation conditions. At each point in time only a very few SPP are excited and also a very few nanoparticles contain non-equilibrium carriers. For example, in the "frame" (a) only one nanoparticle A is excited with SPP and few other nanoparticles tinted red contain non-equilibrium carriers generated by previously excited and decayed SPPs, while majority of nanoparticles tinted yellow remain "cold" or unexcited. In frame (b) the SPP on nanoparticle A has decayed into hot electron-hole pair and no SPPs are present. In frame (c) a new SPP is excited on nanoparticle B while the existing non-equilibrium carriers continue to further cool down

## *2.b Four excitation mechanisms*

We now review the four mechanisms that lead to the decay of SPP.[43] *The first to be mentioned i*s the direct (i.e no phonons or impurities are involved) interband absorption between the inner (4d or 5d ) and outer (5s or 6s) shells of noble metals as shown in Fig 3.a. The energy gap separating the d shell and Fermi level residing in s-shell $E_{ds}$ is equal to 2eV for Au and 3eV for Ag therefore kinetic energy of the electron generated in the s-band (relative to the Fermi level) is only $E_{ib} < \hbar\omega - (E_F - E_d)$ hence only UV excitation can create carriers energetic enough to surpass the barrier $\Phi$ that is on the order of 0.5-1eV. This is shown in Fig.4a where the probability of energy distribution $F_{hot,ib}(E) = 1/(\hbar\omega - E_{ds})$. As far as the angular distribution of the non-equilibrium carriers generated via interband absorption is concerned, it is uniform $R_{ib}(\theta) = 1/2$. as shown in Fig. 5 (a).

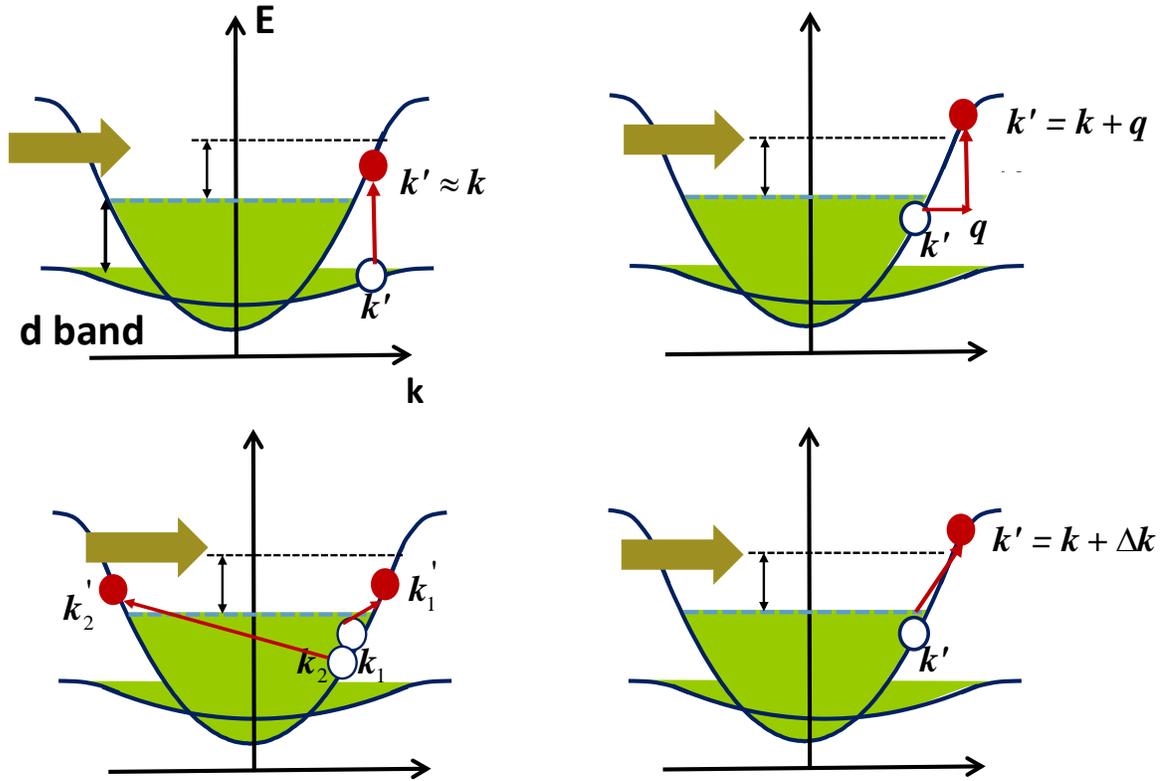

**Fig.3** Four mechanisms of electron hole pairs generation in metals where the average kinetic energy of electrons and holes is low (a) direct "vertical" interband transition (b) phonon (or impurity) assisted transition where the average energy of electrons or holes is ħω/2. (c) Electron-electron Umklapp scattering assisted transition with two electron-hole pairs generated for each SPP and their average kinetic energies being ħω/4. (d) Landau damping or surface collision assisted "tilted" transition where the average energy of electrons or holes is ħω/2. As one can see the carriers generated via processes (b) and (d) are far more likely to have energy sufficient to overcome the surface (Schottky) barrier Φ and be injected into the adjacent semiconductor or dielectric.

As for the holes generated in d-shell, they may have large potential energy relative to the Fermi level, but their kinetic energy is very low, and, more important, their ballistic velocity is very low and hence mean free path is also very short because of their large effective masses. The holes generated inside the metal have very slim chance to reach the surface. Only when the photo-excited (first generation) hole decays into three new second generation particles (two holes and one electron) in s-shell the injection becomes possible, but due to their reduced energies second generation carriers (see below) the probability of such emission is quite low. For this reason, interband absorption only reduces the efficiency of hot carriers.

All other mechanisms are *intraband*, i.e. they involve absorption between two states with different wavevectors in the same s-band. This wave-vector (momentum) mismatch needs to be somehow compensated. *In the second mechanism*, the compensation is provided by either a phonon or an impurity (defect) with wavevector $q$ as shown in Fig. 3b. As a result, as SPP is absorbed and a hot electron and hot hole, each with an average energy of $\hbar\omega/2$ are generated. The energy distribution of the "first generation" of hot carriers is $F_{hot,ph}(E) = 1/\hbar\omega$ where $E_F < E < E_F + \hbar\omega$ for electrons and $E_F > E > E_F - \hbar\omega$ for hot holes as shown in Fig. 4 (b) Conceptually, this process is not different from what is commonly referred to as "Drude" absorption and the SPP damping rate due to this process can be found as $\gamma_{ph}(\omega) = \langle \tau_{ep}^{-1}(E) \rangle_E$ where the electron-phonon (or defect) scattering rate is averaged from $E_F - \hbar\omega$ to $E_F + \hbar\omega$ (while in Drude formula the scattering is taken on Fermi level). The scattering rate for Ag is about $\gamma_{ph} \approx 3 \times 10^{13} s^{-1}$ and for Au $\gamma_{ph} \approx 10^{14} s^{-1}$ [52]. But it is important to stress, that this is a quantum, not a classical process. There is no so-called "classical" [32] or "resistive" [31, 41] contribution to the absorption in which many carriers supposedly instantly created. The energy of the absorbed SPP is almost entirely transferred to just two hot particles –electron and hole, and is not dissipated to a bath of multiple carriers near the Fermi level. That process takes much longer as discussed in the next section. Still, some classical analogies remain true even in the quantum picture – since classically the carriers are accelerated along the direction of optical field, one would expect the photoexcited hot electrons and holes to preferentially travel in that direction. Indeed, detailed calculations in show that photoexcited carriers have normalized angular distribution

$$R_{ph}(\theta) = \frac{3}{4}\cos^2\theta + \frac{1}{4} \qquad (1)$$

as shown in Fig.5, curve b .Since the electric field in most plasmonic structures is close to being normal to the surface, the fraction of hot carriers going towards the surface is twice as large as for uniform distribution.

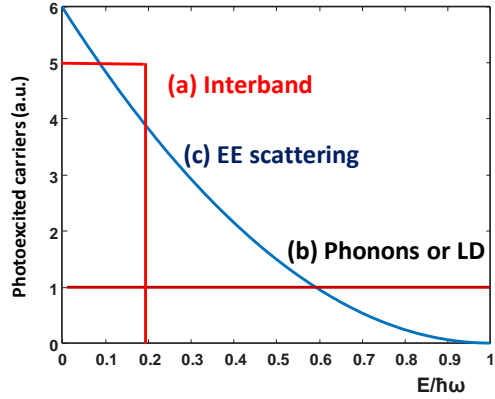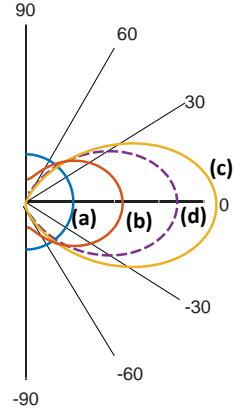

**Fig.4** Energy distributions of the carriers excited via different SPP decay mechanisms (a) –interband transitions (b) phonon/defect assisted transitions or Landau damping (c) electron-electrons scattering assisted transitions. Note that the area under curve c is twice as large as for the other two curves because two electron-hole pairs are generated for each decayed SPP.

**Fig.5** angular distributions of the carriers excited via different SPP decay mechanisms (a) –interband transitions and electron-electrons scattering assisted transitions. (b) phonon/defect assisted transitions (c) Landau damping(d) Effective distribution combining carriers generated by Landau damping on the surface and the carriers generated in the bulk by phonon/defect assisted transitions that have reached the surface

*The third mechanism* by which SPP decays involves the electron-electron (EE) scattering [53, 54](Fig.3c). In this process two electrons and two holes share the energy of the decayed SPP, so on average the energy of each carrier is just $\hbar\omega/4$, so calling them "hot" may not be correct and perhaps "warm" would be a better term. It is well known that at low frequencies EE scattering contribution to the electrical resistance (and therefore ohmic loss) is negligibly small. The reason for that is the fact total momentum of carriers undergoing EE scattering is conserved, i.e. $k_1^{'} + k_2^{'} = k_1 + k_2$. Since as long as the band can be considered parabolic near Fermi surface, the total current can be found as $I = -ev_1/l - ev_2/l = -(e\hbar/lm)(k_1 + k_2)$ where $l$ is the length, $v$ is velocity and $m$ is the effective mass. Therefore, the total current is conserved and no energy is dissipated via EE scattering. But for the optical frequencies the situation is dramatically different since photon energy is sufficiently large to initiate the Umklapp processes [55, 56] in which one of the photoexcited electrons is promoted into the adjacent Brillouin zone so that momentum conservation relation becomes $k_1^{'} + k_2^{'} = k_1 + k_2 + g$ where $g$ the reciprocal lattice vector. Obviously the velocity and current now change as a result of EE scattering and the SPP decays. The EE-scattering assisted SPP damping rate has been found as $\gamma_{ee}(\omega) = F_U(\omega)\tau_{ee}^{-1}(\omega)$ where the EE scattering rate is [57]

$$\tau_{ee}^{-1} \approx \frac{\pi}{24} \frac{E_F}{\hbar} \left(\frac{\hbar\omega}{E_F}\right)^2 \tag{2}$$

and $F_U(\omega)$ is the fraction of the total EE scattering events that are Umklapp processes. This fraction is typically on the order of 0.2-0.5. It follows that EE assisted SPP decay becomes prominent at short wavelengths and for photon energies larger than 2eV $\gamma_{ee} \sim 10^{14} s^{-1}$ i.e. at least as large as phonon assisted SPP damping rate. At the same time for the photon energies less than 1eV, for example for telecom range ($\hbar\omega = 0.8eV$) the EE-assisted damping is not important. The energy distribution of the "first generation" carriers excited with assistance of EE-scattering is

$$F_{hot,ee} = 2 \times 3(\hbar\omega - E_F)^2 / (\hbar\omega)^3, \tag{3}$$

where the factor of 2 indicates that 2 hot electrons are excited by each SPP decay event. This energy distribution is plotted in Fig. 4c. As far as the angular distribution goes, due to involvement of reciprocal vectors the distribution is roughly uniform as shown in Fig. 5a. For all three SPP decay mechanisms outlined so far the spatial distribution of non-equilibrium carrier generation simply follows the density of the SPP energy $\mathcal{E}(\boldsymbol{r})^2$ as shown in Fig.6.

*The fourth and last* SPP decay channel (Fig.3d) is referred either classically (or phenomenologically), as surface collision assisted decay, or in quantum picture as Landau damping [58-61]. Classically, when the electron collides with the surface the momentum can be transferred between the electron and the entire metal lattice, in a way similar to what happens when electron collides with a phonon or defect. That relaxes momentum conservation rules and as first done by Kreibig and Vollmer[62], one can simply introduce the surface collision rate $\gamma_{sc} \sim v_F/d$ where $d$ is the size of nanoparticle. Quantum mechanically, the absorption is the result of the spatial localization of optical field. Since the field is localized its Fourier transform contains all the spectral components, some of them higher than $\Delta k = \omega/v_F$, where $v_F$ is Fermi velocity, which for Au and Ag is about $1.4 \times 10^8 cm/s$. These spectral components provide necessary momentum matching which allow absorption of SPP without assistance from the phonons or defects. This process is commonly referred to as Landau damping[46, 63, 64] and is characterized by the existence of the imaginary part of the wavevector-dependent (nonlocal) dielectric permittivity of the metal described by Lindhard's formula[65]

$$\varepsilon(\omega,k) = \varepsilon_b + \frac{3\omega_p^2}{k^2 v_F^2}\left[1 - \frac{\omega}{2kv_F}\ln\frac{\omega+kv_F}{\omega-kv_F}\right] \tag{4}$$

which obviously has imaginary part for $k > \omega/v_F$. The rate of SPP decay due to Landau damping is

$$\gamma_{LD} = \frac{3}{8}v_F/d_{eff} \tag{5}$$

where

$$d_{eff} = \frac{\int\limits_{metal} \mathcal{E}(\mathbf{r})^2 dV}{\int\limits_{\substack{metal\\surface}} \mathcal{E}_\perp^2(\mathbf{r})dS}, \tag{6}$$

is the volume-to-surface ratio of the mode in the metal and $\mathcal{E}_\perp(\mathbf{r})$ is the normal component of electric field. Clearly both phenomenological and more exact full quantum treatments provide similar results- if one uses (5) and (6) on spherical nanoparticle one obtains $\gamma_{LD} = 0.75 v_F / d$ instead of $\gamma_{sc} = v_F / d$.

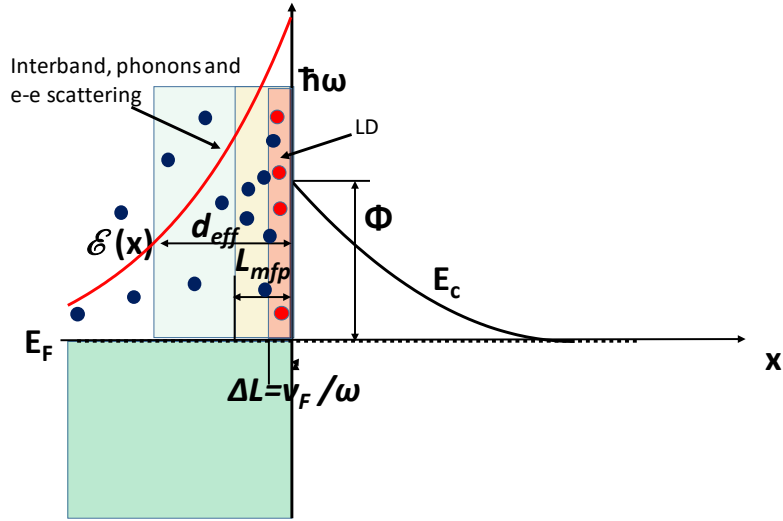

**Fig.6** Energies and locations of non-equilibrium carriers generated via different SPP decay mechanisms. The carriers generated via interband, phonon/defect assisted, and electron-electrons scattering assisted transitions within skin depth (or for small nanoparticle within entire bulk) as shown by black circles. With Landau damping the carriers (red circles) are generated with a layer near the surface ΔL=$v_F$/ω which is much thinner than skin depth.

It is important to note that all the hot carriers generated by surface collisions (Landau damping) are all located within a thin layer of thickness $\Delta L = 2\pi / \Delta k = v_F / v$ where $v$ is an optical frequency shown in Fig.6. As one can see $\Delta L$ is the distance covered by the electron over one optical period, e.g. for gold and 700nm excitation is only about 3nm, which is obviously shorter than mean free path of electron between collisions (typically 10-20nm). Therefore, one half of the carriers excited via Landau damping will always end up at the surface, which is one reason why Landau damping is the most favorable mechanism of carrier generation for their injection from the metal. The second reason is that the angular distribution of the carriers excited via LD is highly nonuniform,

$$R_{LD}(\theta) \sim 2|\cos^3\theta|, \tag{7}$$

as shown in Fig.5c. As one can see the fraction of carriers that impinges on the surface at normal incidence is increased by a factor of 4 compared to the uniform distribution and by a factor of 2 compared to the distribution of the carriers generated by phonon-assisted processes.

## 3. Numbers: Hot carriers decay, but how fast?

*3.a Is there such thing as equilibrium electron temperature and how hot it can be?*

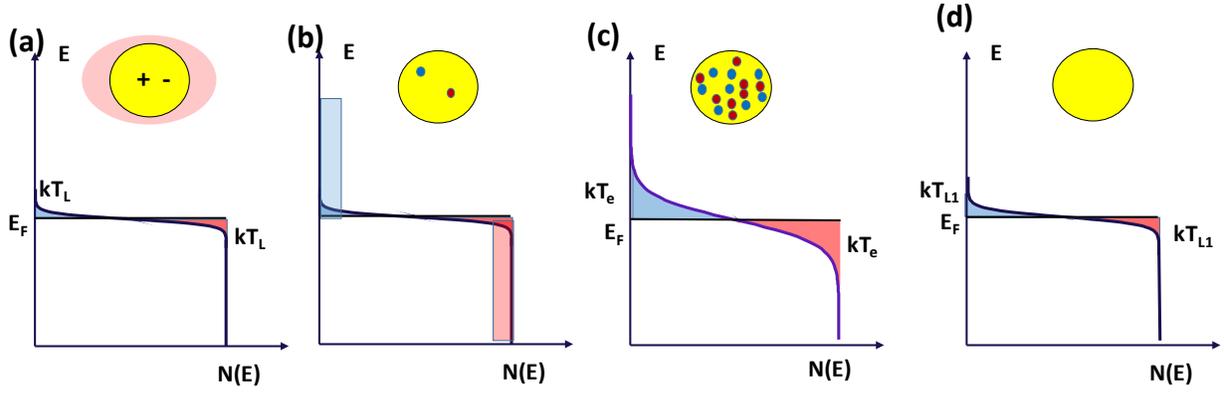

**Fig.7** Conventionally assumed picture of hot carrier generation and relaxation in metal nanoparticles. (a) An SPP is excited on a nanoparticle while the carriers inside are distributed according to Fermi Dirac statistics with equilibrium lattice temperature T$_L$ (b) After the SPP lifetime τ$_{SPP}$ this SPP has decayed and as a result non-equilibrium electron-hole pairs with energies ranging from E$_F$-hω to E$_F$+hω. (c) After the electron thermalization time commensurate with τ$_{ee}$ all the electrons are distributed according to Fermi-Dirac statistics with electron temperature T$_e$> T$_L$ (d) After the electron-lattice thermalization time τ$_{el}$ the electron and lattice are at equilibrium with a new lattice temperature T$_{L1}$ > T$_L$ . This temperature will eventually decrease back to equilibrium lattice temperature T$_L$ after time τ$_L$ determined by the nanoparticle environment.

So we have established that SPP shown in Fig. 7a decays within its lifetime $\tau_{SPP} \sim \gamma^{-1}$ where $\gamma$ the sum of all four SPP decays rates is outlined above a well as radiative decay rate $\gamma_{rad}$ that is typically small in comparison. Once the "first generation" non-equilibrium carriers have been excited as shown in Fig.7b they decay via both electron-electron and phonon (or defect) assisted processes. Conventionally, it is assumed that fast electron scattering $\tau_{ee} \sim 10\,fs$ [49, 50, 66-68] defined in (2) quickly establishes thermal equilibrium among the electrons with temperature $T_e$ that is significantly higher than the equilibrium lattice temperature $T_L$ (Fig.7c) Then electron temperature relaxes as the energy is transferred to the lattice with characteristic relaxation time $\tau_{el}$ that is a couple of orders of magnitude longer than $\tau_{ee}$ as shown in Fig. 7d. Often this time is erroneously referred to as electron-phonon scattering time, but this is of course wrong – the electron phonon scattering time $\tau_{ep}$, defined in *previous section* as is roughly the same order of magnitude as $\tau_{ee}$ but it takes many scattering events to reduce the energy of hot electrons because in each event only a small amount of energy is being lost by an electron. It is for this reason that first generation carriers that underwent a phonon scattering event can be considered quasi-ballistic since they largely keep their energy and their distribution in the momentum space does not change significantly.

The whole process in general cannot be characterized with two separate relaxation times, especially since $\tau_{ee}$ gradually increases as electrons lose energy, while the $\tau_{el}$ gradually decreases as there are more secondary hot carriers are generated. We shall return to it later, but at any rate it is reasonable to assume that for some time all the energy absorbed by the metal as a result of SPP decay is stored in the electron gas causing effective rise of average temperature

$$\overline{T}_e - T_L \sim \frac{\gamma_{nr}\tau_{el}}{C_{el}} u_{SPP} = \frac{\gamma_{nr}\tau_{el}}{C_{el}} F^2 I_{IN} n/c \tag{8}$$

where specific heat of electrons can roughly be estimated as

$$C_{el} = \frac{\pi^2}{2} \frac{k_B^2 T}{E_F} N_e \approx 0.025 k_B N_e \approx 0.018 J/K \cdot cm^3 \tag{9}$$

where $E_F = 5.56 eV$ (for Au) is Fermi energy and $N_e = 5.9 \times 10^{22} cm^{-3}$ (for Au) is the density of free electrons. The specific heat is low because only about 1 out of 40 carriers residing within $\pi^2 k_B T/4 \approx 64 meV$ Fermi level participate in heat exchange. Substituting (9) into (8) and assuming realistic value of $\gamma_{nr}\tau_{el} \sim 100$ we obtain

$$\overline{T}_e - T_L \sim 2 \times 10^{-7} F^2 I_{IN} \ K, \tag{10}$$

where input irradiance is once again in units of W/cm². With $F^2 = 10^3$ and input irradiance of 1000 suns (100W/cm²) the average electron temperature rise is miniscule at 0.02K. But what about the instant rise? According to our estimates the average time interval between two SPP absorption events is typically longer than the "storage time" $\tau_{el}$. The temperature rise following absorption of a single SPP is simply

$$\Delta T_{e,inst} = \hbar\omega / C_{el} V \approx 1.6 \times 10^4 \ K \cdot nm^3 / V \tag{11}$$

The instant increase is much higher than the average one and for really small nanoparticle of 10nm diameter it can be as high as 12K which is easy to understand as the energy of the first generation electron-hole pair is eventually split between roughly 700 thermally active electrons residing near Fermi level. But even with this temperature increase (which in unit of energy corresponds to about 1meV) there would be no noticeable impact on carriers surmounting the energy barrier $\Phi$ that is typically hundreds of meV high. *One should note that the result (11) is remarkable as it points to the quantum nature of hot carrier generation with the equilibrium temperature of hot carriers depending only on the volume of nanoparticle and the photon energy and totally independent on input power density!* Furthermore, even if the excitation is continuous, the picture of all carriers settling at the same average electron temperature $\overline{T}_e$ is deeply flawed – the correct picture is that at any given time, a relatively small fraction of nanoparticles have electrons in them excited to some temperatures $\Delta T_{e,inst}$ which are different for different nanoparticles due to their different volumes, while the majority of nanoparticles experience no electron temperature increase at all as shown Fig. 1 where the vast majority of the nanoparticles remain "cold" at any given moment. Also, it is important to note that once the energy is transferred to the lattice the temperature of the lattice $T_{L1}$ is different from the equilibrium $T_L$ and it takes for some time $\tau_L$ to reach equilibrium. This time $\tau_L$ is determine mostly by how efficient is the energy transfer between the nanoparticle and the surrounding (or adjacent) medium. If $\tau_L$ is longer than the mean interval between the SPP excitations of the same nanoparticle $\tau_L$ two the equilibrium lattice temperature $T_L$ increases resulting in thermionic emission over the barrier.

Overall, the intermediary conclusion here is that if one posits that the hot carriers that have thermalized at some electron temperature $T_e$, these carriers definitely do not have sufficient energy to contribute to the injection into semiconductor or dielectric and most probably they cannot contribute to chemical reaction on the metal surface either, as stated by Sivan[39, 40]. (The reason for ambiguity here is that the nature of the energy barrier that the electron must surpass to cause chemical reaction is not clearly defined.) Therefore, one must follow the decay of excited hot carriers step-by-step in order to ascertain the probability of them being injected out of the metal after each step, where the step is defined by a single collision of hot carriers with the thermal carriers (EE scattering) or with phonons and impurities.

## *3.b Generations of nonequilibrium carriers*

These relaxation steps are shown in Fig. 8. On the left the SPP decay causes excitation of a single electron-hole pair – the first generation with energy $E_{1,n}$ and $\sum_{n=1}^{2} E_{1,n} = \hbar\omega$ (assuming that the holes energies are counted down from the Fermi level). Then either electron or hole scatter off the electron residing below the Fermi level thus creating another electron-hole pair. Once both electron and hole scatter once (which on average should take time $\tau_{ee}$) there are three second generation holes and electrons each with energies $E_{2,n}$, and $\sum_{n=1}^{6} E_{2,n} = \hbar\omega$. After on average another time interval $\tau_{ee}$ elapses (this interval may be longer than original since the EE scattering is energy dependent) each of the second generation carriers engenders 3 third generation carriers with energies $E_{3,n}$, and $\sum_{n=1}^{18} E_{2,n} = \hbar\omega$. The process continues until the average energy of the M-th generation, $\langle E_{M,n} \rangle_n = \hbar\omega / 2 \cdot 3^{M-1}$ becomes comparable to $k_B T$ so that there is no distinction between the "hot" and "cool" carriers, therefore it takes roughly the time

$$\tau_{e,cool} = (M-1)\tau_{ee} \approx \log_3(\hbar\omega / 2k_B T)\tau_{ee} \tag{12}$$

for the electrons to cool down to some kind of equilibrium between themselves. For the SPP energies below 1.5 eV it takes no more than 3 scattering events to cool down the electrons hence $\tau_{e,cool} < 4\tau_{ee}$.

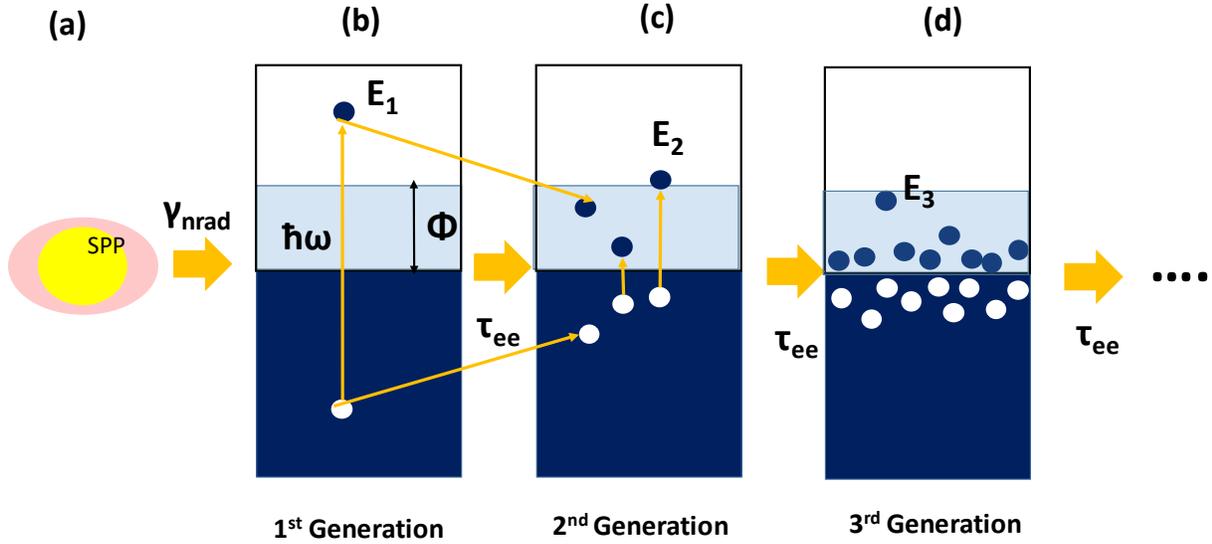

**Fig.8** Quantum picture of carriers generation and relaxation in metal nanoparticles. (a) An SPP is excited on a nanoparticle (b) An SPP has decayed engendering a primary (1st generation) electron-hole pair (c) Each of the 1st generation carriers decays into 3 second generation carriers (d) Each of the 2nd generation carriers decays into 3 third generation carriers.

So, while cool-down time is of the same order of magnitude as EE-scattering time, it is definitely larger than it by a factor of a few, as the ubiquitous statement stubbornly permeating the literature that a single scattering event is sufficient to establish the equilibrium of the electrons [31, 32, 36, 37, 41] is incorrect. Obviously, during this time interval there will be electron-phonon scattering events, because, remember, that $\tau_{ee}$ and $\tau_{ep}$ are roughly of the same order of magnitude. However, these events cause insignificant loss of energy for each hot carrier and thus can be safely disregarded.

Let us now consider the distribution of the second through fourth generation hot carriers in energy space. When the electron of the first generation hot carriers with energy distribution $f_1(E) = \delta(E - E_1)$, decays into three new second generation carriers their distribution is

$$f_2(E, E_1) = 2(E_1 - E) / E_1^2 \qquad (13)$$

Then these carriers decay into 9 third generation carriers, whose distribution can be found as

$$f_3(E, E_1) = \int_E^{E_1} f_2(E_2, E_1) f_2(E, E_2) dE_2 = \frac{4}{E_1^2} \left[ (E_1 + E) \log \frac{E_1}{E_3} - 2(E_1 - E) \right] \qquad (14)$$

and then into 27 carriers of the fourth generation becomes

$$f_4(E, E_1) = \int_{E4}^{E_1} f_3(E_3, E_1) f_2(E, E_3) dE_3 = \frac{4}{E_1^2} \left[ (E_1 - E_4) \left( \log^2 \frac{E}{E_4} + 12 \right) - 6(E_1 + E_4) \log \frac{E}{E_4} \right] \qquad (15)$$

Note that even though the functions $f_m$ for m>2 diverge near zero energy, they are all perfectly integrable to $\int_0^{E_1} f_m(E,E_1)dE = 1$ The distributions of total number of carriers in each generation, $N_m(E,E_1) = 3^{m-1} f_m(E,E_1)$ are shown in Fig. 9. with energies $E_m$ scaled relative to energy $E_1$. As one can see, the distribution quickly shifts to lower energies, however, when plotted on log scale in Fig. 9b the curves are not linear and therefore one cannot ascribe a single electron temperature $T_e$ to the carriers.

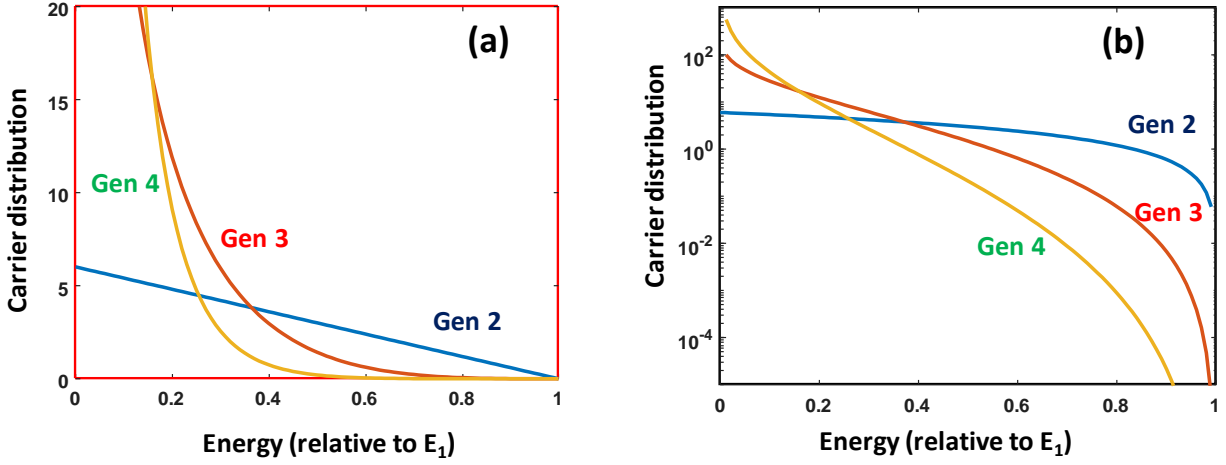

**Fig.9** Energy distribution of the 2nd through 4th generation of carriers generated as a result of decay of a single first generation carrier with energy $E_1$. (a) linear scale (b) logarithmic scale.

Next we determine the distribution of all carriers generated by photons with energy $\hbar\omega$ as

$$f_m(E,\hbar\omega) = \int_{E_m}^{\hbar\omega} f_m(E,E_1)dE_1 \qquad (16)$$

and obtain (assuming original distribution associated with phonon, defect assisted, or LD process)

$$f_1(E,\hbar\omega) = \frac{1}{\hbar\omega}; \quad E \leq \hbar\omega$$

$$f_2(E,\hbar\omega) = \frac{2}{\hbar\omega}\left(\frac{E}{\hbar\omega} - 1 - \log\frac{E}{\hbar\omega}\right)$$

$$f_3(E,\hbar\omega) = \frac{2}{\hbar\omega}\left(\log^2\left(\frac{E}{\hbar\omega}\right) + 2\log\frac{E}{\hbar\omega}\left(\frac{E}{\hbar\omega} + 2\right) + 6\left(1 - \frac{E}{\hbar\omega}\right)\right)$$

$$f_4(E,\hbar\omega) = \frac{4}{\hbar\omega}\left(-\frac{1}{3}\log^3\left(\frac{E}{\hbar\omega}\right) + \log^2\left(\frac{E}{\hbar\omega}\right)\left(\frac{E}{\hbar\omega} - 3\right) - 4\log\frac{E}{\hbar\omega}\left(2\frac{E}{\hbar\omega} + 3\right) - 20\left(1 - \frac{E}{\hbar\omega}\right)\right)$$

(17)

The carrier number distributions $N_m(E,\hbar\omega) = 3^{m-1} f_m(E,\hbar\omega)$ for the first four generations of carriers are plotted in Fig. 10 a and b As one can see, within roughly time $\tau_{e,cool} \sim 3\tau_{ee}$ the distribution changes dramatically and in fact resembles the distribution one would expect if one used classical Drude model in which absorption light generates many low energy carriers via "friction" , but it is important that in quantum picture this does not happen instantly and hot carriers may depart the metal before they decay. Also, even for the fourth generation of carriers one cannot introduce equilibrium temperature $T_e$ as evident from the figure 10b where the negative slope of distribution increases at higher energies indicating reduced number of high energy carriers capable of surpassing energy barrier.

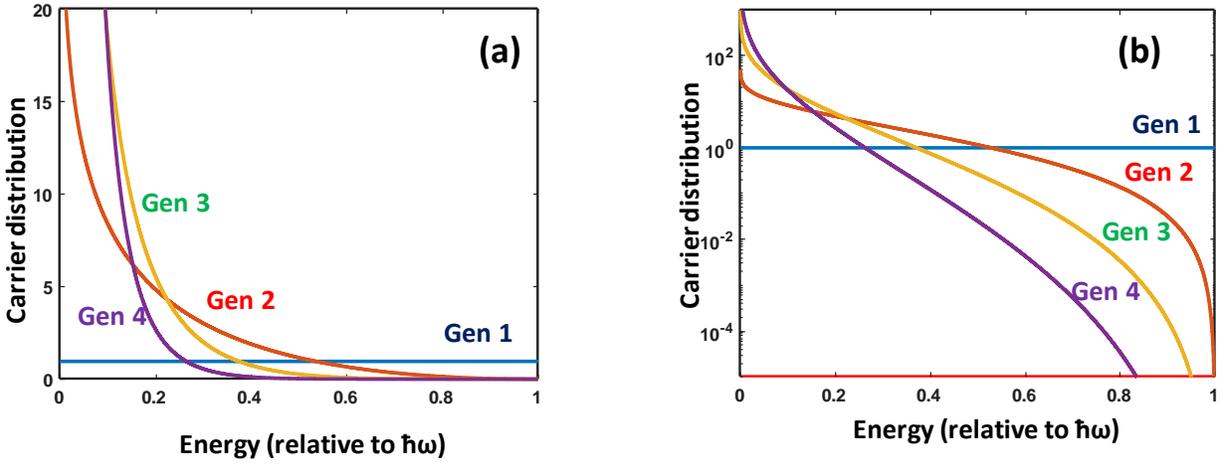

**Fig.10** Energy distribution of the 1st through 4th generation of carriers generated as a result of decay of a single SPP with energy **ℏω** (a) linear scale (b) logarithmic scale. Note that the distribution cannot be defined by a single electron temperature $T_e$.

Let us now estimate what are the chances for the hot carriers of each generation to overcome a potential barrier $\Phi$. Two cases will be considered. In the first case we assume that the transverse momentum is conserved and the efficiency of carrier extraction is [6, 7]

$$\eta_{ext,m}^{(+)}(\Phi,\omega) \sim \frac{m_s}{2m_m E_F} \int_{\Phi}^{\hbar\omega} (E-\Phi) f_m(E,\hbar\omega) dE \tag{18}$$

where $m_s$ and $m_m$ are the effective masses of metal The results are shown in Fig. 11(a) (without the term in front of the integral as we are only interested in the relative strength of the injection of carriers from different generations) . As one can see, the probability of extraction decreases dramatically in each generation for the barrier height that is at least 30% of the photon energy.

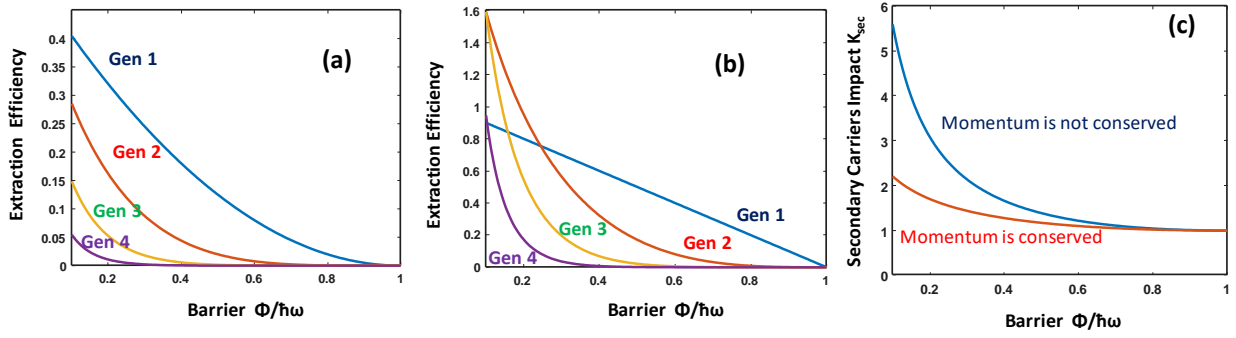

**Fig.11** Extraction efficiency of the primary and secondary generations of nonequilibrium carriers of different generations as a function of barrier height. (a) momentum conservation is enforced (b) momentum conservation is violated (c) Impact of the secondary carriers.

For the second case we assume that the momentum conservation rules are fully relaxed and therefore all we need is to evaluate the total number of the carriers with energy above the barrier [33]

$$\eta^{(-)}_{ext,m}(\Phi,\omega) \sim C \int_{\Phi}^{\hbar\omega} f_m(E,\hbar\omega)dE \qquad (19)$$

where C is band structure dependent factor to be derived in the next section, but at this point we are only interested in relative impact of the secondary electrons and holes.) The results are shown in Fig.11 b and are similar to the case of complete momentum conservation, although the secondary carriers become important for the barriers that are less than half of the photon energy. We summarize the impact of extraction of secondary electrons in Fig.11 c where we plot the function $K_{sec} = \sum_{m=1}^{4} \eta_{ext,m}/\eta_{ext,1}$. As one can see for the barrier that is at least half as large as phonon energy the impact of secondary electrons is negligible no matter what model we assume.

For the infrared detectors, which is one of the more promising hot carrier applications, it is desirable to have barrier height relatively to reduce thermal noise, and, as shown in maximum detectivity is achieved at $\hbar\omega - \Phi \approx 4k_B T \ll \hbar\omega$. Hence the impact of secondary carriers can be completely neglected – for all practical purposes once a single EE scattering event takes place the carriers a no longer capable of overcoming the barrier. For other cases where the barrier is relatively low one can simply use the semi-empirical expression to modify the time it takes the primary (first generation) carriers to decay to the point where they no longer overcome the barrier as $\tau_{ee}^{eff} = K_{sec}\tau_{ee}$ with $K_{sc}$ typically less than 2.

These results are also relevant to the carriers generated via EE-assisted and interband absorption. The holes generated via interband absorption in the d-shall can decay into two holes and one electron, all in the s-band, where they can move relatively fast, and some of those carriers may have energy sufficient to exit across the barrier. The energy distribution of these second generation carriers is similar to Fig.9 for $E_1 = (E_F - E_d)$. As long as the barrier is close to photon energy one can completely neglect the injection of intraband-absorption generated carriers, otherwise one can simply add their relatively small contribution to $K_{sec}$. At any rate, once interband absorption commences, the Q of the SPP mode decreases and so does the field enhancement, thus negating the whole goal plasmonic assisted detection or catalysis. Similarly, judging from Fig. 4 one cannot expect a large contribution from the carriers generated with EE scattering help. Once again, that contribution can definitely be ignored for infrared light and for visible light that contribution can also be incorporated into $K_{\text{sec}}$.

So, to conclude this section, we state that for all practical purposes only the primary (first generation) carriers generated with the phonon/defect assistance or via Landau damping are the ones that can find their way out of the metal. Once these carriers undergo a single EE scattering event, their energies will for the most part be way too small to overcome the barrier on the metal/semiconductor(dielectric) interface. We shall refer to these carriers as "quasi-ballistic" since these carriers are expected to propagate quasi-ballistically (phonon and defect scattering does not reduce energy significantly) towards the interface and then get injected across the barrier. The distribution into which the secondary electrons created as result of EE scattering eventually settle cannot be characterized by a single electron temperature $T_e$ and for practical values of the incident light intensity never contribute to the injection over a reasonably high barrier. It is harder to speculate whether this conclusion also holds for the process of photocatalysis on the surface of the metal, as these processes are not yet entirely understood. Still, for a reasonably high activation energy it seems that only the quasi-ballistic carriers have sufficient energy to initiate the chemical reaction.

## 4. Exodus. Hot carriers are injected from metal into the semiconductor or dielectric. How efficiently?

### 4.a Transport efficiency

Let us now establish the efficiency of the hot electron injection, $\eta_{ext}(\Phi,\omega) = N_{ext}/N_{SPP}$ where $N_{ext}$ is the number of carriers injected into semiconductor/dielectric. This efficiency can be split into two factors, the transport efficiency $\eta_{tran}(\omega) = N_s/N_{SPP}$ where $N_s$ is the number of carriers reaching the surface of the metal, and the extraction efficiency $\eta_{ext}(\Phi,\omega) = N_{inj}/N_{s.}$.

To estimate the transport efficiency we first introduce the mean free path of hot carriers, $L_{mfp,e} = v_F \tau_{ee}^{eff}$. Note that this definition is different from the mean free path distribution in the Drude transport theory as it involves only EE scattering, since, as mentioned above, the collisions with phonons or defects do not affect the energy of hot carriers. Then we can introduce the "surface proximity factor"

$$\Gamma_{prox} = \frac{\int\limits_{metal} \mathcal{E}(\mathbf{r})^2 \left\langle \exp(-R/L_{mfp,e}) \right\rangle_\Omega dV}{\int\limits_{metal} \mathcal{E}(\mathbf{r})^2 dV} \leq 1 \qquad (20)$$

where R is the distance to the surface and averaging is done over the solid angle. Obviously for small nanoparticles with dimension less than $L_{mfp}$ the proximity factor approaches unity. With that the transport efficiency becomes

$$\eta_{tran} = \frac{\gamma_{LD} + \Gamma_{prox}\gamma_{ph}}{\gamma_{ib} + \gamma_{ee} + \gamma_{LD} + \gamma_{ph} + \gamma_{rad}}, \tag{21}$$

where $\gamma_{rad}$ is the radiative decay rate, which is quite small for nanoparticles with radius less than 50nm as well as for the propagating SPP mode. We assume that operates below the offset of interband absorption and with insignificant EE scattering one can obtain the expression for the spherical nanoparticle

For the case of propagating SPP in which the intensity inside the metal decays exponentially as $\exp(-x/L_p)$ one can estimate $\Gamma_{prox} = L_{mfp,e}/(L_p + L_{mfp,e})$ where $L_p$ is the penetration depth and one obtains

$$\eta_{tran} = \frac{1}{2}\frac{\frac{3}{8}\frac{v_F}{L_p} + \frac{L_{mfp,e}}{L_p + L_{mfp,e}}\frac{1}{\tau_{ep}}}{\frac{3}{8}\frac{v_F}{L_p} + \frac{1}{\tau_{ep}}} = \frac{1}{2} - \frac{L_p^2}{2(L_p + L_{mfp,e})(L_p + \frac{3}{8}L_{mfp,p})} \tag{22}$$

where mean free path due to phonon and defect collisions is $L_{mfp,p} = v_F\tau_{ep}$ and the factor ½ accounts for the fact that only a half of hot carriers move towards the surface For the Au or Ag guide in the near IR range where the Drude approximation for the dielectric constant $\varepsilon_m(\lambda) \approx 1 - \lambda^2/\lambda_p^2$ the penetration depth $L_p \approx \lambda_p/4\pi \approx 12nm$ where $\lambda_p = 140nm$ is plasma wavelength. For $\tau_{ee}^{eff} = 10fs$ and $\tau_{ep} = 15fs$ we obtain $L_{mfp,e} = 14nm$, $L_{mfp,p} = 40nm$, and $\eta_{tran} \approx 40\%$. For the case of small spherical nanoparticle with diameter d we obtain $\Gamma_{prox} \approx 0.7\exp(-d/2L_{mfp,e})$ and

$$\eta_{tran} = \frac{\frac{3}{8}\frac{v_F}{d} + 0.7\exp(-d/2L_{mfp,e})\frac{1}{\tau_{ep}}}{\frac{3}{8}\frac{v_F}{d} + \frac{1}{\tau_{ep}}} = \frac{1 + 1.84\exp(-d/2L_{mfp,e})d/L_{mfp,p}}{1 + 2.66d/L_{mfp,p}} \tag{23}$$

where we have neglected the small possibility of the LD carriers generated at one end of a nanoparticle going all the way to the other end without catering. For d=20nm nanosphere one gets $\eta_{tran} \approx 52\%$ and for d=40nm $\eta_{tran} \approx 30\%$. Since the carriers generated by EE and phonons have different angular distributions the overall distribution of carriers near the surface is

$$R_{eff}(\theta) \approx a_{ph}R_{ph}(\theta) + (1-a_{ph})R_{LD}(\theta) = a_{ph}\left(\frac{3}{4}\cos^2\theta + \frac{1}{4}\right) + 2(1-a_{ph})|\cos^3\theta| \tag{24}$$

where $a_{ph}$ is the fraction of the carriers generated via phonon/defect scattering and according to (22) for the propagating SPP $a_{ph} \approx 0.45$, while according to (24) for the spherical nanoparticles with $12nm < d < 60nm$ $a_{ph}$ stays between 0.4 and 0.45. In other words, phonon and defect assisted absorption is responsible for almost one half of the ballistic carriers arriving at the metal surface, and one can write for the effective angular distribution as $R_{eff}(\theta) \approx 1.1|\cos^3\theta| + 0.33\cos^2\theta + 0.11$ that is shown in Fig. 5d

## *4.b Extraction efficiency*

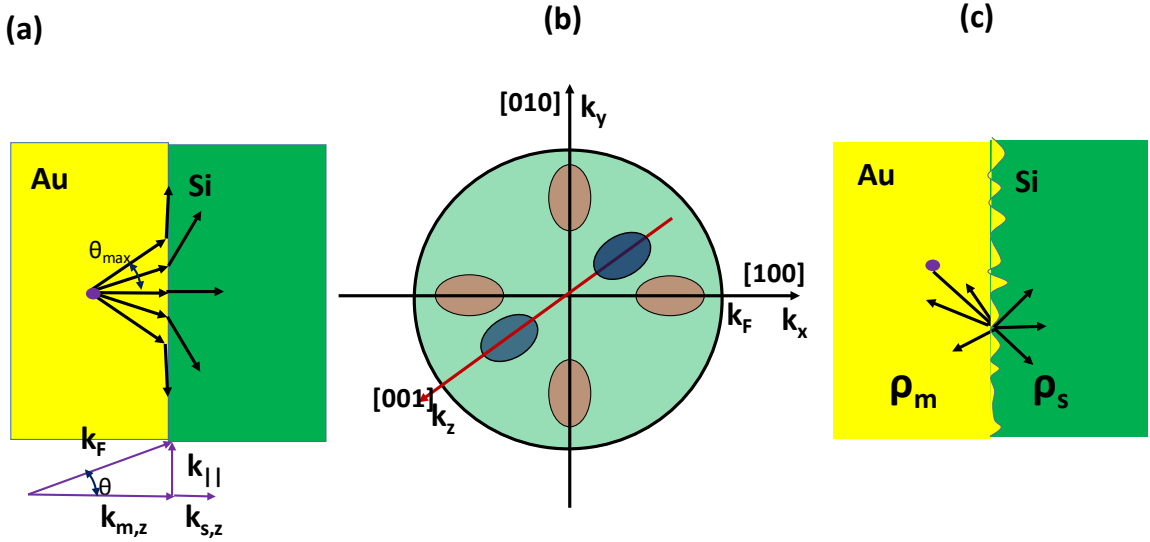

**Fig.12 (**a) Carrier injection from the metal into silicon across a smooth interface. Only the electrons with incidence angle less than θ$_{max}$ get injected. (b) 6 valleys in the conduction band of Si. Electrons get injected only into the two valley along [100] direction. (c) Carrier injection from the metal into silicon across a rough interface with momentum conservation restriction entirely lifted.

With transport to the surface out of the way, we consider the extraction efficiency of all the quasi-ballistic carriers arriving at the surface, i.e. transmission coefficient over the barrier $\Phi$. As shown in Fig.12a, If the lateral (in plane) wavevector is continuous across the barrier, i.e. $\mathbf{k}_{m,\|} = \mathbf{k}_{s,\|} = \mathbf{k}_\|$ then the longitudinal wavevectors for the electron in the metal whose energy above Fermi level is $E$ and whose incidence angle is $\theta$ can be found as $k_{m,z} = \sqrt{2m_m/\hbar^2\left(E + E_F - k_\|^2\right)} \approx k_F \cos\theta$, where $m_m$ is the effective mass of metal. For the semiconductor $k_{s,z} \approx k_F\sqrt{\sin^2\theta_{max} - \sin^2\theta}$, where $\theta_{max}(E,\Phi) = \sin^{-1}\sqrt{(m_s/m_m)(E-\Phi)/E_F}$ and $m_s$ is the effective mass of semiconductor. As a result of wavevector continuity (lateral momentum conservation) only the carriers with $\theta \leq \theta_{max}(E,\Phi)$ can be extracted from the metal. First we assume that all the carriers within the "extraction cone" can exit the metal, hence

$$\eta_{ext}(\hbar\omega, \Phi) = \int_0^{\hbar\omega} f_1(E,\hbar\omega) d\hbar\omega \int_0^{\theta_{max}(E,\Phi)} R_{eff}(\theta) \sin\theta d\theta dE \qquad (25)$$

where $f_1(E, \hbar\omega) = 1/\hbar\omega$ according to (17). In our calculations we consider injection from a noble metal, like Au into Si. The conduction band of Si, characterized by 6 valleys along <100> directions, each characterized by longitudinal $m_L = 0.98 m_0$ and transverse $m_T = 0.19 m_0$ as shown in Fig. 12b. The barrier is treated as variable parameter due to presence of surface states. The effective injection is possible almost exclusively into two valleys along [100] (normal to the interface) direction since in the other four valleys the Bloch functions have symmetry that is almost orthogonal to S-states in the metal. We then approximate Si conduction band with a single valley isotropic band with effective density of state mass of $m_s = 2^{2/3} m_L^{1/2} m_T^{2/3} = 0.52 m_0$. If the injection is into the valence band, the effective mass is almost exactly the same, $m_s = 0.49 m_0$.

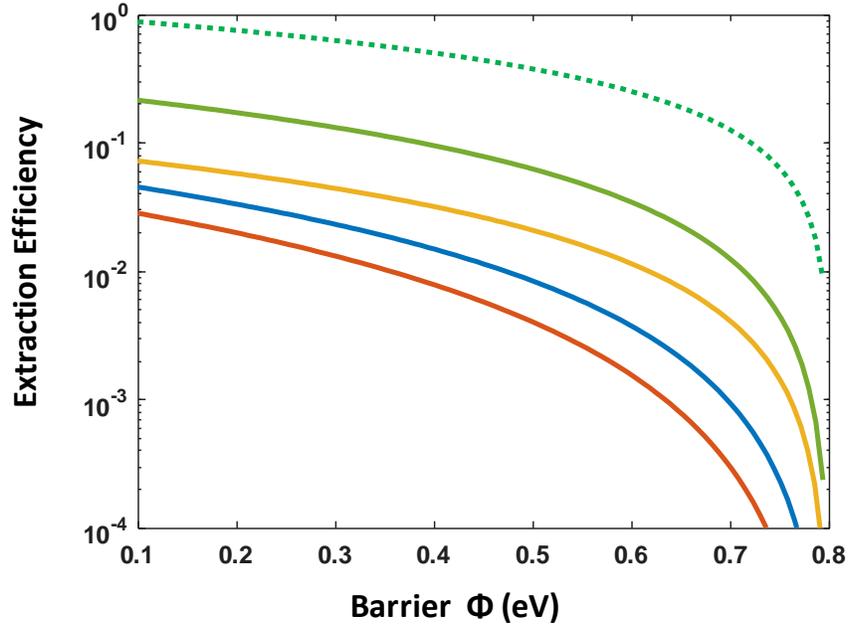

**Fig.13** Extraction efficiency of the hot carriers generated on the Metal/Si interface by the SPPs excited with λ=1500nm photons. (a) Smooth Interface, reflection is not taken into account (Eq.(27)) (b) Smooth Interface, reflection is taken into account (Eq.(27))(c) Rough interface Complete extraction of all the above-the-barrier carriers (Eq. (28)) (d) Rough interface, momentum conservation rules are relaxed but the transitions into transverse valleys of Si are not allowed (Eq.(29)) (e) Rough interface, momentum conservation rules are relaxed and the transitions into transverse valleys of Si are allowed (Eq.(29)).

Then, since typically $\theta_{max}$ is rather small, we can neglect the angular dependence of ballistic carriers distribution assuming $R_{eff}(\theta) \approx R_{eff}(0)$ and obtain

$$\eta_{ext}(\hbar\omega, \Phi) \approx \frac{1}{4} R_{eff}(0) \frac{m_s}{m_0} \frac{(\hbar\omega - \Phi)^2}{\hbar\omega E_F} \qquad (26)$$

or essentially a Fowler's formula [69], as plotted in Fig.13a for $\hbar\omega = 0.8 eV$ (($\lambda = 1500 nm$) and is practically no different from the exact formula (25).

Next we shall take into consideration the reflection from the interface and include transmission coefficient

$$\eta_{ext}(\hbar\omega,\Phi) = \int_0^{\hbar\omega} f_1(E,\hbar\omega)d\hbar\omega \int_0^{\theta_{max}(E,\Phi)} \left[1 - \left(\frac{k_{m,z}/m_0 - k_{m,s}/m_s}{k_{m,z}/m_0 + k_{m,s}/m_s}\right)^2\right] R_{eff}(\theta)\sin\theta d\theta dE \quad (27)$$

into the integral in (25) – the result is shown in Fig.13b and as one can see extraction efficiency is reduced by about 20% for low barriers but by as much as a factor of 2 for barrier that is only 200meV below the photon frequency.

As one can see, lateral momentum conservation severely restricts the "exit cone" of the incident ballistic carriers to the small angle $\theta_{max} < \pi/10$ for the visible and even les for near IR wavelengths leading to small extraction efficiencies. However, experimental data show that higher injection efficiencies can be achieved when the momentum conservation is no longer valid due to extreme disorder at the interface. In [33] the injection efficiency of nearly 30% for Au/GaAs interface was reported, while high efficiencies for injection into TiO$_2$ from Au nanoparticles has been measured in [34]. Increase in photocurrent in the photodetectors with rough Au/Si interface relative to the ones with a smooth interface has been reported in [].

To explain these extraordinary results, the simplest model was proposed in [33] that assumed that all the hot carriers with energies higher than barrier Φ can be extracted as in (19), leading to

$$\eta_{ext,\max}(\hbar\omega,\Phi) = \frac{\hbar\omega - \Phi}{\Phi} \quad (28)$$

Shown in Fig.13c but this approach entirely neglects the possibility of backscattering into the metal. Another approach [70] was to use the explicit interface roughness scattering explicitly to obtain the enhancement of extraction efficiency by a factor of a few. That model, however, could only be applied to a relatively small roughness, and, as matter of fact, neglected enhanced backscattering as well.

To find the ultimate extraction efficiency we shall follow the theory of Yablonovitch[44] developed for the seemingly different task of light trapping in the dielectric with roughened surface. Essentially the argument developed there can be applied to the case of the surface roughened to the degree that momentum conservation is no longer valid – then according to Fermi golden rule the rate of scattering in a given direction depends only on the density of states as shown in Fig.12c Now, if the densities of states in the metal and semiconductor are $\rho_m$ and $\rho_s$ respectively, one can obtain a rather simple expression for the extraction efficiency

$$\eta_{ext,\max}(\hbar\omega,\Phi) = \int_0^{\hbar\omega} f_1(E,\hbar\omega)\frac{\rho_s(E)}{\rho_s(E)+\rho_m(E)}dE = \frac{1}{\hbar\omega}\int_0^{\hbar\omega}\frac{(m_s/m_0)^{3/2}(E-\Phi)^{1/2}/E_F^{1/2}}{(m_s/m_0)^{3/2}(E-\Phi)^{1/2}/E_F^{1/2}+1}dE$$
(29)

For small extraction probability one can obtain the estimate

$$\eta_{ext}(\hbar\omega,\Phi) = \frac{2}{3}\left(\frac{m_s}{m_0}\right)^{3/2}\frac{(\hbar\omega-\Phi)^{3/2}}{\hbar\omega E_F^{1/2}} \quad (30)$$

As one can see the result (30) is not nearly as high as the estimate (28) (by a factor of roughly $(m_s/m_0)^{3/2}(\hbar\omega/\Phi-1)^{1/2}$ ) yet it is higher than the extraction efficiency without roughness (26) by roughly a factor $(m_0/m_s)^{1/2}(\hbar\omega/\Phi-1)^{1/2}$. For Si it is important to consider what is effective density of state mass to use in (29) and (30). If one assumes that the electron can only be injected into two X-valleys, one should use previously defined $m_s = 0.52m_0$ effective mass. However, it is possible for the disorder to be so strong that it allows transitions into the other four valleys which would imply density of states mass equal to $m_s = 1.08m_0$ which will increase the extraction efficiency by a factor of three. Both of the curves are plotted in Fig.1d and e respectively. Whether one can induce such a strong disorder that it will break the selection rules that prevent the injection into the "transverse valleys" is difficult to state, but it is definitely not inconceivable. Therefore, if one assumes that the barrier height is 0.4eV, one can see that introducing roughness can increase the extraction efficiency by a factor between 4 and 12, from less than 1% to more than 10% leading to overall efficiency of 4-5%. The increase is even more dramatic for higher barriers, which is where the performance of detector becomes optimal due to decrease of dark current.

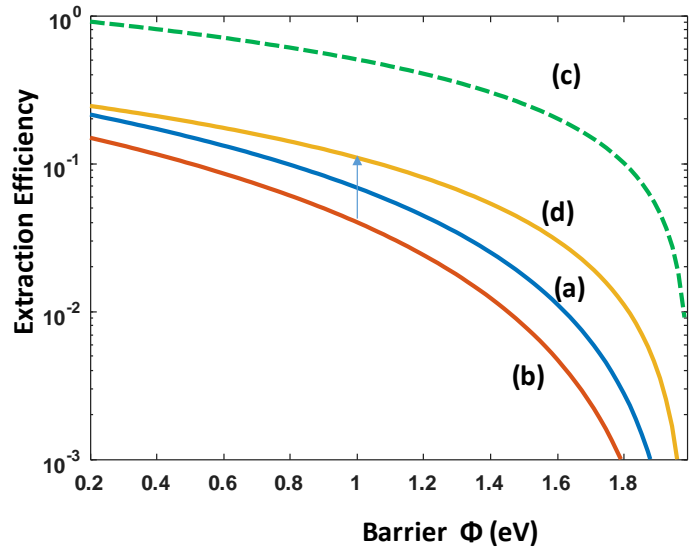

**Fig.14** Extraction efficiency of the hot carriers generated on the Metal/TiO$_2$ interface by the SPPs excited with λ=620nm photons. (a) Smooth Interface, reflection is not taken into account (Eq.(26)) (b) Smooth Interface, reflection is taken into account (Eq.(27)) (c) Rough interface Complete extraction of all the above-the-barrier carriers (Eq.(28)) (d) Rough interface, momentum conservation rules are relaxed (Eq (30))

In Fig. 14 we show the extraction efficiency with and without disorder for the case of Au/TiO$_2$ interface and wavelength of 620 nm (photon energy of 2eV). The effective mass of TiO$_2$ is $m_s = 0.8m_0$ and for relatively high barrier one gets significant improvement of extraction efficiency due to scattering on rough interface. But typically the barrier height on metal-TiO2 interface is only a few hundred meV and roughness only increases $\eta_{ext}$ by a factor of 2 or so. Overall, one can expect extraction efficiencies on the scale of 10-20%. Thus the fact that with TiO$_2$ one attains higher injection efficiency can be traced to the fact that it has large density of states available for injection.

Overall, the injection efficiency for Au/TiO$_2$ can be as high as 10% or even higher if one assumes that the carriers can travel from one side of nanoparticle to the other. For Au/Si the number is smaller, primarily due to smaller density of states and higher barrier, and is typically on the scale of a few percent. So, while quantum efficiency of hot carrier detectors can never reach those easily attainable in commercial photodiodes, it may be sufficient for these detectors to find some applications because they can absorb long wave radiation while being compatible with commercial silicon technology. Whether the same can be said about plasmon-assisted phtocatalysis remains to be seen,

## 5. Conclusions

At this point it is time to summarize the main points made in the present work so the busy reader can be spared the arduous task of going through all the derivations and calculations performed in prior sections and instead focus on the practical consequences of this work.

First important point made in this work relates to the *inherently discrete, quantum nature of SPP generation and decay.* Under realistic illumination, with optical powers densities far less than a MW/cm$^2$ at any given time no more than a single SPP is present on a typical nanoparticle ( same is true for a number of SPPs inside a plasmonic photodetector waveguide when the optical power is less than 10 $\mu W$ ). Consequently, at any given time, the number of non-equilibrium carriers in a given nanoparticle (or inside a waveguide) is typically only a few and their combined energy is exactly $\hbar\omega$. Therefore, one cannot use "average" electron temperature T$_e$ to describe evolution of hot carriers as their distribution is never thermal. The one and only way to describe evolution of hot carriers is to simply follow them through the scattering events, generation after generation.

The second point is that as long as the height of barrier separating the metal from the adjacent isolator/semiconductor is at least moderately large (higher than $\hbar\omega/3$) only *two SPP decay mechanisms out of four: Landau damping and phonon/defect assisted decay generate significant number carriers with energies high enough to surpass the barriers.* Furthermore, after only a single event of electron-electron scattering a sufficiently energetic "first generation" carrier decays into three carriers whose energies are too low for the extraction across the barrier. Thus hot carriers have only a very short (~10 fs) time over which they can be injected from metal into semiconductor/isolator. After that no injection can take place. For this reason, the carriers generated at the surface via Landau damping stand the highest chance of being injected.

The *third point* is that the extraction efficiency of hot carriers is *greatly affected by the smoothness* of the interface. For smooth interface the momentum conservation dictates that only a very small fraction of hot carriers with small in-plane wavevectors are capable of exiting the metal. Typical injection efficiencies do not exceed 0.1%. But if the surface roughness is so high that momentum is no longer conserved, practically all the carriers with energies above the barrier stand a chance of exiting this metal. Hower, this chance is far from 100% because the density of states near Fermi in the metal is typically much higher than in semiconductor. Still, depending on the barrier height and effective mass, up to 10% injection efficiencies are possible. Whether the momentum is conserved or not, it is desirable to have semiconductor/dielectric with large effective mass (and thus density of states).

*Finally,* when it comes to photodetectors, the main practical parameter is detectivity and for that one must reduce thermal noise- hence one does not benefit from lowering the barrier beyond the optimal 4kT. Choosing a semiconductor with large density of states, such as Si or Ge rather than, say, GaAs remains to be the only viable stratagem for the performance improvement in addition to the aforementioned roughening of the interface and engineering the waveguide mode to make sure that the field is concentrated near the surface and Landau damping is a dominant SPP decay channel. For the photocatalysis one should also increase the relative strength of Landau damping by using smaller nanoparticles with large surface to volume ratio and roughen the interface, and, since thermal noise is not a factor, the barrier should be lowered. However, when the barrier is sufficiently low, most of the enhancement of catalysis will come not from hot carriers per se but simply by the thermionic emission due to the increase of the ambient temperature- and heating can be achieved by means other than light absorption.

*In the end*, I have presented here a compilation of factors determining efficiency of plasmonic assisted hot carrier injection for applications in (mostly) detectors and (also) photocatalysis. Some of the results presented here have been of course investigated before, as, for instance, the fact that nonequilibrium carrier distribution is not thermal have been argued by many. But other results are indeed entirely new. In particular, the important fact that for realistic illumination conditions no more than a single SPP gets excited per nanoparticle, surprisingly, has been overlooked before. Also, the paramount role of density of states in the injection process has not been given proper attention. It is my belief that it is valuable for the plasmonic community to combine in one place coherent and unified description of all the steps of the process –from SPPs generation through their decay engendering non-equilibrium carriers that then go through competing processes of decay, transport to the surface and extraction from metal. This treatment provides a simple way of estimating overall efficiency of the injection and also outlines the pathways to its optimization. Whether research community finds any value in this modest effort is an open question, but I sure hope that not far in the future hot carrier devices will enter the mainstream and this work will play a helpful role, no matter how small, in it.

**Acknowledgment**: Support of NSF grant Grant # 1507749 has been a key to writing this article. Aside from that, valuable if sometimes heated discussions with Prof. P. Noir as well as points made by the new member of his research group Ms. S. Artois have been essential to this work.

**Figure Captions**

**Fig.1** Geometry of hot electron-hole pair generation in the metal with subsequent injection into semiconductor/dielectric using (a) Localized SPP's (mostly used in photocatalysis) and (b) Propagating SPPs (mostly used in photodetection). At realistic optical powers SPPs never coexist with hot carriers at the same time.

**Fig.2** Dynamics of hot carrier generation and decay in the array of metal nanoparticles under typical photo-excitation conditions. At each point in time only a very few SPP are excited and also a very few nanoparticles contain non-equilibrium carriers. For example, in the "frame" (a) only one nanoparticle A is excited with SPP and few other nanoparticles tinted red contain non-equilibrium carriers generated by previously excited and decayed SPPs, while majority of nanoparticles tinted yellow remain "cold" or unexcited. In frame (b) the SPP on nanoparticle A has decayed into hot electron-hole pair and no SPPs are present. In frame (c) a new SPP is excited on nanoparticle B while the existing non-equilibrium carriers continue to further cool down.

**Fig.3** Four mechanisms of electron hole pairs generation in metals where the average kinetic energy of electrons and holes is low (a) direct "vertical" interband transition (b) phonon (or impurity) assisted transition where the average energy of electrons or holes is $\hbar\omega/2$. (c) Electron-electron Umklapp scattering assisted transition with two electron-hole pairs generated for each SPP and their average kinetic energies being $\hbar\omega/4$. (d) Landau damping or surface collision assisted "tilted" transition where the average energy of electrons or holes is $\hbar\omega/2$. As one can see the carriers generated via processes (b) and (d) are far more likely to have energy sufficient to overcome the surface (Schottky) barrier Φ and be injected into the adjacent semiconductor or dielectric

**Fig.4** Energy distributions of the carriers excited via different SPP decay mechanisms (a) –interband transitions (b) phonon/defect assisted transitions or Landau damping (c) electron-electrons scattering assisted transitions. Note that the area under curve c is twice as large as for the other two curves because two electron-hole pairs are generated for each decayed SPP.

**Fig.5** angular distributions of the carriers excited via different SPP decay mechanisms (a) –interband transitions and electron-electrons scattering assisted transitions. (b) phonon/defect assisted transitions (c) Landau damping(d) Effective distribution combining carriers generated by Landau damping on the surface and the carriers generated in the bulk by phonon/defect assisted transitions that have reached the surface

**Fig.6** Energies and locations of non-equilibrium carriers generated via different SPP decay mechanisms. The carriers generated via interband, phonon/defect assisted, and electron-electrons scattering assisted transitions within skin depth (or for small nanoparticle within entire bulk) as shown by black circles. With Landau damping the carriers (red circles) are generated with a layer near the surface $\Delta L = v_F/\omega$ which is much thinner than skin depth.

**Fig.7** Conventionally assumed picture of hot carrier generation and relaxation in metal nanoparticles. (a) An SPP is excited on a nanoparticle while the carriers inside are distributed according to Fermi Dirac statistics with equilibrium lattice temperature $T_L$ (b) After the SPP lifetime $\tau_{SPP}$ this SPP has decayed and as a result non-equilibrium electron-hole pairs with energies ranging from $E_F-\hbar\omega$ to $E_F+\hbar\omega$. (c) After the electron thermalization time commensurate with $\tau_{ee}$ all the electrons are distributed according to Fermi-Dirac statistics with electron temperature $T_e > T_L$ (d) After the electron-lattice thermalization time $\tau_{el}$ the electron and lattice are at equilibrium with a new lattice temperature $T_{L1} > T_L$. This temperature will eventually decrease back to equilibrium lattice temperature $T_L$ after time $\tau_L$ determined by the nanoparticle environment.

**Fig.8** Quantum picture of carriers generation and relaxation in metal nanoparticles. (a) An SPP is excited on a nanoparticle (b) An SPP has decayed engendering a primary (1st generation) electron-hole pair (c) Each of the 1st generation carriers decays into 3 second generation carriers (d) Each of the 2nd generation carriers decays into 3 third generation carrier

**Fig.9** Energy distribution of the 2nd through 4th generation of carriers generated as a result of decay of a single first generation carrier with energy $E_1$. (a) linear scale (b) logarithmic scale.

**Fig.10** Energy distribution of the 1st through 4th generation of carriers generated as a result of decay of a single SPP with energy $\hbar\omega$ (a) linear scale (b) logarithmic scale. Note that the distribution cannot be defined by a single electron temperature $T_e$.

**Fig.11** Extraction efficiency of the primary and secondary generations of nonequilibrium carriers of different generations as a function of barrier height. (a) momentum conservation is enforced (b) momentum conservation is violated (c) Impact of the secondary carriers.

**Fig.12 (**a) Carrier injection from the metal into silicon across a smooth interface. Only the electrons with incidence angle less than $\theta_{max}$ get injected. (b) 6 valleys in the conduction band of Si. Electrons get injected only into the two valley along [100] direction. (c) Carrier injection from the metal into silicon across a rough interface with momentum conservation restriction entirely lifted.

**Fig.13** Extraction efficiency of the hot carriers generated on the Metal/Si interface by the SPPs excited with λ=1500nm photons. (a) Smooth Interface, reflection is not taken into account (Eq.(27)) (b) Smooth Interface, reflection is taken into account (Eq.(27))(c) Rough interface Complete extraction of all the above-the-barrier carriers (Eq. (28)) (d) Rough interface, momentum conservation rules are relaxed but the transitions into transverse valleys of Si are not allowed (Eq.(29)) (e) Rough interface, momentum conservation rules are relaxed and the transitions into transverse valleys of Si are allowed (Eq.(29)).

**Fig.14** Extraction efficiency of the hot carriers generated on the Metal/$TiO_2$ interface by the SPPs excited with λ=620nm photons. (a) Smooth Interface, reflection is not taken into account (Eq.(26)) (b) Smooth Interface, reflection is taken into account (Eq.(27)) (c) Rough interface Complete extraction of all the above-the-barrier carriers (Eq.(28)) (d) Rough interface, momentum conservation rules are relaxed (Eq (30))